\documentclass[10pt,a4paper,onecolumn]{article}
\usepackage{marginnote}
\usepackage{graphicx}
\usepackage{xcolor}
\usepackage{authblk,etoolbox}
\usepackage{titlesec}
\usepackage{calc}
\usepackage{tikz}
\usepackage{hyperref}
\hypersetup{colorlinks,breaklinks=true,
            urlcolor=[rgb]{0.0, 0.5, 1.0},
            linkcolor=[rgb]{0.0, 0.5, 1.0}}
\usepackage{caption}
\usepackage{tcolorbox}
\usepackage{amssymb,amsmath}
\usepackage{ifxetex,ifluatex}
\usepackage{seqsplit}
\usepackage{xstring}

\usepackage{float}
\let\origfigure\figure
\let\endorigfigure\endfigure
\renewenvironment{figure}[1][2] {
    \expandafter\origfigure\expandafter[H]
} {
    \endorigfigure
}

\usepackage{fixltx2e} 
\usepackage[
  backend=biber,
]{biblatex}
\bibliography{paper.bib}

\let\textttOrig=\texttt
\def\texttt#1{\expandafter\textttOrig{\seqsplit{#1}}}
\renewcommand{\seqinsert}{\ifmmode
  \allowbreak
  \else\penalty6000\hspace{0pt plus 0.02em}\fi}

\makeatletter
\let\href@Orig=\href
\def\href@Urllike#1#2{\href@Orig{#1}{\begingroup
    \def\Url@String{#2}\Url@FormatString
    \endgroup}}
\def\href@Notdoi#1#2{\def\tempa{#1}\def\tempb{#2}%
  \ifx\tempa\tempb\relax\href@Urllike{#1}{#2}\else
  \href@Orig{#1}{#2}\fi}
\def\href#1#2{%
  \IfBeginWith{#1}{https://doi.org}%
  {\href@Urllike{#1}{#2}}{\href@Notdoi{#1}{#2}}}
\makeatother

\newlength{\cslhangindent}
\newlength{\csllabelwidth}
\newenvironment{CSLReferences}[3] 
 {
  \setlength{\parindent}{0pt}
  \ifodd #1 \everypar{\setlength{\hangindent}{\cslhangindent}}\ignorespaces\fi
  \ifnum #2 > 0
  \setlength{\parskip}{#2\baselineskip}
  \fi
 }%
 {}
\usepackage{calc}

\usepackage[top=3.5cm, bottom=3cm, right=1.5cm, left=1.0cm,
            headheight=2.2cm, reversemp, includemp, marginparwidth=4.5cm]{geometry}

\titleformat{\section}
  {\normalfont\sffamily\Large\bfseries}
  {}{0pt}{}
\titleformat{\subsection}
  {\normalfont\sffamily\large\bfseries}
  {}{0pt}{}
\titleformat{\subsubsection}
  {\normalfont\sffamily\bfseries}
  {}{0pt}{}
\titleformat*{\paragraph}
  {\sffamily\normalsize}

\usepackage{fancyhdr}
\makeatletter
\let\ps@plain\ps@fancy
\fancyheadoffset[L]{4.5cm}
\fancyfootoffset[L]{4.5cm}

\definecolor{linky}{rgb}{0.0, 0.5, 1.0}

\newtcolorbox{repobox}
   {colback=red, colframe=red!75!black,
     boxrule=0.5pt, arc=2pt, left=6pt, right=6pt, top=3pt, bottom=3pt}

\newcommand{\ExternalLink}{%
   \tikz[x=1.2ex, y=1.2ex, baseline=-0.05ex]{%
       \begin{scope}[x=1ex, y=1ex]
           \clip (-0.1,-0.1)
               --++ (-0, 1.2)
               --++ (0.6, 0)
               --++ (0, -0.6)
               --++ (0.6, 0)
               --++ (0, -1);
           \path[draw,
               line width = 0.5,
               rounded corners=0.5]
               (0,0) rectangle (1,1);
       \end{scope}
       \path[draw, line width = 0.5] (0.5, 0.5)
           -- (1, 1);
       \path[draw, line width = 0.5] (0.6, 1)
           -- (1, 1) -- (1, 0.6);
       }
   }

\patchcmd{\@maketitle}{center}{flushleft}{}{}
\patchcmd{\@maketitle}{center}{flushleft}{}{}
\patchcmd{\@maketitle}{\LARGE}{\LARGE\sffamily}{}{}
\def\maketitle{{%
  
  \AB@maketitle}}
\makeatletter
\renewcommand\AB@affilsepx{ \protect\Affilfont}
\renewcommand\AB@affilnote[1]{{\bfseries #1}\hspace{3pt}}
\renewcommand{\affil}[2][]%
   {\newaffiltrue\let\AB@blk@and\AB@pand
      \if\relax#1\relax\def\AB@note{\AB@thenote}\else\def\AB@note{#1}%
        \setcounter{Maxaffil}{0}\fi
        \begingroup
        \let\href=\href@Orig
        \let\texttt=\textttOrig
        \let\protect\@unexpandable@protect
        \def\thanks{\protect\thanks}\def\footnote{\protect\footnote}%
        \@temptokena=\expandafter{\AB@authors}%
        {\def\\{\protect\\\protect\Affilfont}\xdef\AB@temp{#2}}%
         \xdef\AB@authors{\the\@temptokena\AB@las\AB@au@str
         \protect\\[\affilsep]\protect\Affilfont\AB@temp}%
         \gdef\AB@las{}\gdef\AB@au@str{}%
        {\def\\{, \ignorespaces}\xdef\AB@temp{#2}}%
        \@temptokena=\expandafter{\AB@affillist}%
        \xdef\AB@affillist{\the\@temptokena \AB@affilsep
          \AB@affilnote{\AB@note}\protect\Affilfont\AB@temp}%
      \endgroup
       \let\AB@affilsep\AB@affilsepx
}
\makeatother

\renewcommand\Affilfont{\sffamily\small\mdseries}
\ifnum 0\ifxetex 1\fi\ifluatex 1\fi=0 
  \usepackage[T1]{fontenc}
  \usepackage[utf8]{inputenc}

\else 
  \ifxetex
    \usepackage{mathspec}
    \usepackage{fontspec}

  \else
    \usepackage{fontspec}
  \fi
  \defaultfontfeatures{Ligatures=TeX,Scale=MatchLowercase}

\fi
\IfFileExists{upquote.sty}{\usepackage{upquote}}{}
\IfFileExists{microtype.sty}{%
\usepackage{microtype}
\UseMicrotypeSet[protrusion]{basicmath} 
}{}

\usepackage{hyperref}
\hypersetup{unicode=true,
            pdftitle={pyssam -- a Python library for statistical modelling of biomedical shape and appearance},
            pdfborder={0 0 0},
            breaklinks=true}
\usepackage{color}
\usepackage{fancyvrb}

\DefineVerbatimEnvironment{Highlighting}{Verbatim}{commandchars=\\\{\}}
\newenvironment{Shaded}{}{}

\newcommand{\BuiltInTok}[1]{#1}

\newcommand{\ControlFlowTok}[1]{\textcolor[rgb]{0.00,0.44,0.13}{\textbf{#1}}}

\newcommand{\DecValTok}[1]{\textcolor[rgb]{0.25,0.63,0.44}{#1}}

\newcommand{\ImportTok}[1]{#1}

\newcommand{\KeywordTok}[1]{\textcolor[rgb]{0.00,0.44,0.13}{\textbf{#1}}}
\newcommand{\NormalTok}[1]{#1}
\newcommand{\OperatorTok}[1]{\textcolor[rgb]{0.40,0.40,0.40}{#1}}

\let\addcontentslineOrig=\addcontentsline
\def\addcontentsline#1#2#3{\bgroup
  \let\texttt=\textttOrig\addcontentslineOrig{#1}{#2}{#3}\egroup}
\let\markbothOrig\markboth
\def\markboth#1#2{\bgroup
  \let\texttt=\textttOrig\markbothOrig{#1}{#2}\egroup}
\let\markrightOrig\markright
\def\markright#1{\bgroup
  \let\texttt=\textttOrig\markrightOrig{#1}\egroup}

\usepackage{graphicx,grffile}
\makeatletter
\def\maxwidth{\ifdim\Gin@nat@width>\linewidth\linewidth\else\Gin@nat@width\fi}
\def\maxheight{\ifdim\Gin@nat@height>\textheight\textheight\else\Gin@nat@height\fi}
\makeatother
\setkeys{Gin}{width=\maxwidth,height=\maxheight,keepaspectratio}
\IfFileExists{parskip.sty}{%
\usepackage{parskip}
}{
\setlength{\parindent}{0pt}
\setlength{\parskip}{6pt plus 2pt minus 1pt}
}
\ifx\paragraph\undefined\else
\let\oldparagraph\paragraph
\renewcommand{\paragraph}[1]{\oldparagraph{#1}\mbox{}}
\fi
\ifx\subparagraph\undefined\else
\let\oldsubparagraph\subparagraph
\renewcommand{\subparagraph}[1]{\oldsubparagraph{#1}\mbox{}}
\fi

\title{\texttt{pyssam} -- a Python library for statistical modelling of
biomedical shape and appearance}

        \author[1]{Josh Williams}
          \author[1]{Ali Ozel}
          \author[1]{Uwe Wolfram}
    
      \affil[1]{School of Engineering and Physical Sciences, Heriot-Watt
University, Edinburgh, UK}
  \date{\vspace{-7ex}}

\begin{document}
\maketitle

\marginpar{

  \begin{flushleft}
  \sffamily\small

  {\bfseries DOI:} \href{https://doi.org/TBD}{\color{linky}{TBD}}

  \vspace{2mm}

  {\bfseries Software}
  \begin{itemize}
    \setlength\itemsep{0em}
    \item \href{https://github.com/jvwilliams23/pyssam}{\color{linky}{Repository}} \ExternalLink
    \item \href{https://doi.org/10.5281/zenodo.7509407}{\color{linky}{Archive}} \ExternalLink
  \end{itemize}

  \vspace{2mm}

  \par\noindent\hrulefill\par

  \vspace{2mm}

  {\bfseries Editor:} Pending Editor \ExternalLink \\
  \vspace{1mm}
    {\bfseries Reviewers:}
  \begin{itemize}
  \setlength\itemsep{0em}
        \item Pending Reviewers
    \end{itemize}
    \vspace{2mm}

  {\bfseries Submitted:} N/A\\
  {\bfseries Published:} N/A

  \vspace{2mm}
  {\bfseries License}\\
  Authors of papers retain copyright and release the work under a Creative Commons Attribution 4.0 International License (\href{http://creativecommons.org/licenses/by/4.0/}{\color{linky}{CC BY 4.0}}).

  \end{flushleft}
}

\hypertarget{summary}{%
\section{Summary}\label{summary}}

\texttt{pyssam} is a Python library for creating statistical shape and
appearance models (SSAMs) for biological (and other) shapes such as
bones, lungs or other organs. A point cloud best describing the
anatomical `landmarks' of the organ are required from each sample in a
small population as an input. Additional information such as landmark
gray-value can be included to incorporate joint correlations of shape
and `appearance' into the model. Our library performs alignment and
scaling of the input data and creates a SSAM based on covariance across
the population. The output SSAM can be used to parameterise and quantify
shape change across a population. \texttt{pyssam} is a small and low
dependency codebase with examples included as Jupyter notebooks for
several common SSAM computations. The given examples can easily be
extended to alternative datasets, and also alternative tasks such as
medical image segmentation by incorporating a SSAM as a constraint for
segmented organs.

\hypertarget{statement-of-need}{%
\section{Statement of need}\label{statement-of-need}}

Statistical shape (and appearance) models (SSAMs) have drawn significant
interest in biomedical engineering and computer vision research due to
their ability to automatically deduce a linear parameterisation of shape
covariances across a small population of training data (Baka et al.,
2011; Cootes et al., 1995; Heimann \& Meinzer, 2009; Väänänen et al.,
2015). The classic statistical shape model (SSM) approach uses a point
cloud of landmarks which are in correspondence across several instances
of a shape. The covariances of how the shape changes across the training
population are computed, and principal component analysis (PCA) is used
to parameterise the different modes of shape variation (Cootes et al.,
1995). This approach paved the way for automatic algorithms which could
significantly aid medical image segmentation (similar to an atlas)
(Irving et al., 2011), characterise how the organ shape varies over a
population as a diagnostic tool (Osanlouy et al., 2020), or even
reconstruct a full 3D structure from a sparser imaging modality such as
planar X-ray images (Baka et al., 2011; Väänänen et al., 2015).

We have found that available open-source toolkits such as Statismo and
Scalismo (Lüthi et al., 2012) suffer from an exhaustive number of
dependencies and are difficult to adapt to new tasks, datasets and I/O
datatypes. ShapeWorks (Cates et al., 2017) is another strongly developed
library for statistical shape modelling, but it uses an alternative
method of extracting landmarks (a so-called particle-based method) which
is less broadly used and more complex than a landmark-based system
(where landmarks can be defined in any desired way for different
anatomical shapes). Additionally, as the machine learning ecosystem has
strong foundations in Python, building statistical models in C++, Scala
or other languages reduces compatibility with the majority of modern
machine learning developments (Bhalodia et al., 2018). We therefore
implemented a lightweight Python framework for SSAMs which is easily
adaptable with few dependencies, making it suitable for integrating as
part of a broader codebase, as well as installing and running on
high-performance computing clusters where users do not have root access
to install many dependencies. We provide Jupyter notebooks on
\href{https://pyssam.readthedocs.io/en/latest/}{readthedocs} and two
example datasets that allow users new to coding or SSAMs to learn how
these models work in an interactive way to ease access when learning a
new research topic and library.

\hypertarget{overview}{%
\section{Overview}\label{overview}}

\begin{figure}
\centering
\includegraphics[width=1\textwidth,height=\textheight]{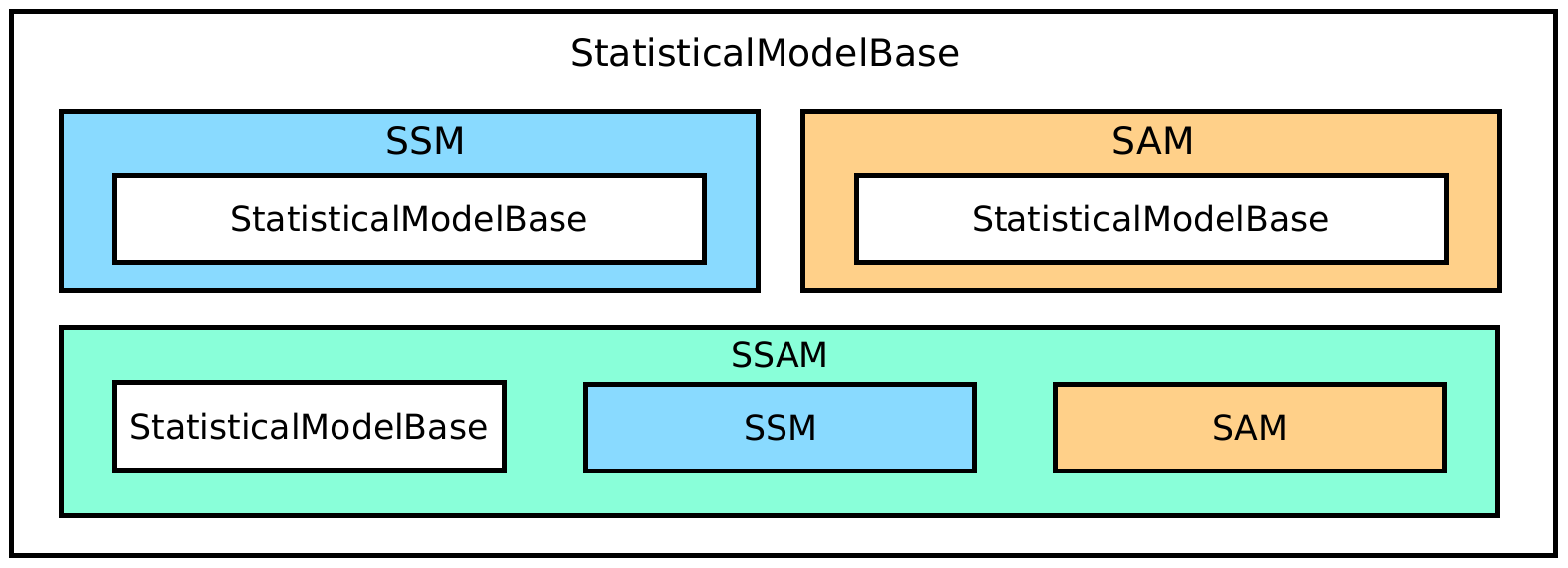}
\caption{Schematic overview of the codebase. Each modelling class is
abstracted from the \texttt{StatisticalModelBase} class and contains
several inherited variables such as model weights and principal
components. The \texttt{SSAM} class inherits from
\texttt{StatisticalModelBase}, but also uses pre-processing pipelines
from \texttt{SSM} and \texttt{SAM}.\label{fig:code}}
\end{figure}

The main modelling classes are built on the abstract base class
\texttt{StatisticalModelBase}, which has several methods for
pre-processing data and performing PCA (\autoref{fig:code}). There are
also several global variables that are inherited which are related to
principal components, component variances and model parameters. The
classes for \texttt{SSM} and \texttt{SAM} pre-process the data (align to
zero mean and standard deviation of one) and can compute the population
mean shape/appearance. Finally, the \texttt{SSAM} class for shape and
appearance modelling inherits all of these, but also imports the
\texttt{SSM} and \texttt{SAM} methods to pre-process shape and
appearance features separately, before they are merged into one dataset
for modelling.

\section{Examples}

Here we present two example applications of \texttt{pyssam}. The first
example examines shape variations in a toy dataset created for this
study, which has a tree structure. Tree structures appear often in
biology, including the lung airways and vascular system. Toy datasets
such as these are a simple means to visualise and interpret the
modelling and code framework. We then provide a more complex example
which considers the left lower lobe of human lungs obtained from CT data
(Tang et al., 2019). This example considers shape and appearance, where
the appearance is the gray-value at the landmark location on an X-ray
projection (obtained with the \texttt{AppearanceFromXray} helper class).

\subsection{Statistical shape modelling toy dataset} \label{sec:tree}

\begin{figure}
\centering
\includegraphics[width=1\textwidth,height=\textheight]{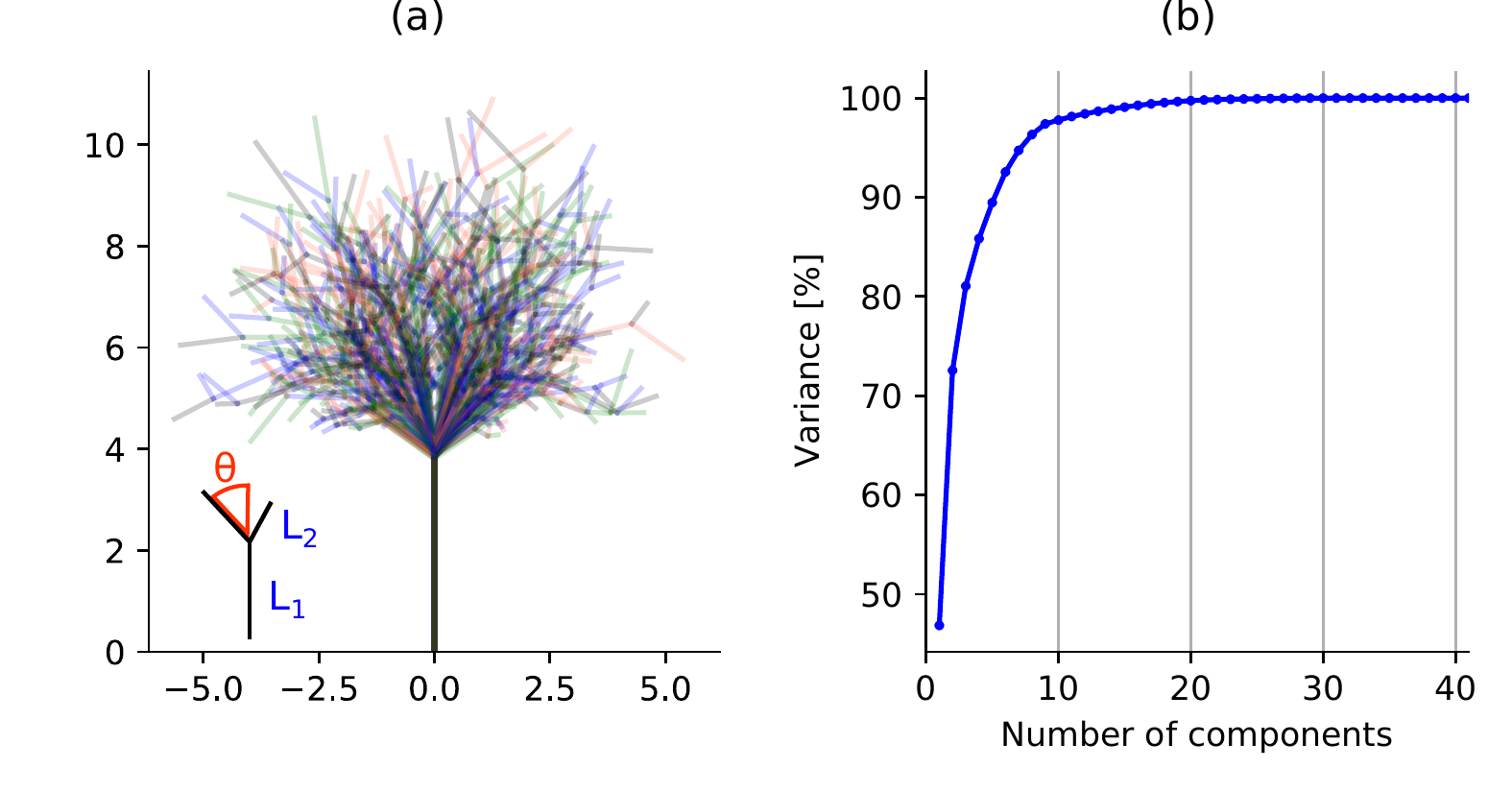}
\caption{Overview of tree dataset population. Panels show (a) a
visualisation of 100 tree samples, and (b) cumulative variance versus
the number of PCA components constructed by the statistical shape model.
Inset of (a) shows a legend describing the morphological parameters
varied to create the tree dataset. These parameters include the initial
branch length, \(L_1\), the branch length ratio \(L_R = L_2/L_1\), and
branching angle \(\theta\).\label{fig:tree}}
\end{figure}

To understand the shape modelling process, we have provided a dataset
class called \texttt{Tree} which creates a number of tree shapes which
are randomly computed based on global minimum and maximum values for
angle and branch length ratio (between parent and child). Tree
parameters are shown in \autoref{fig:tree}a. Tree nodes are converted to
a numpy array and used to initialise \texttt{pyssam.SSM}. At
initialisation of the \texttt{SSM} class, the landmarks are aligned,
scaled to unit standard deviation and stacked into a matrix of shape
\((N_f, 3N_L)\) where \(N_f\) is the number of features (samples in our
training dataset) and \(N_L\) is the number of landmarks (each with a
\(x,y,z\) coordinates). All \(y\) coordinates in this case are zero,
meaning the data is actually 2D but we preserve a 3D coordinate system
for simplicity in generalising the code to more common 3D applications.
The code below shows how we can simply obtain a SSM from a set of
landmarks.

\begin{Shaded}
\begin{Highlighting}[]
\ImportTok{from}\NormalTok{ glob }\ImportTok{import}\NormalTok{ glob}
\ImportTok{import}\NormalTok{ numpy }\ImportTok{as}\NormalTok{ np}
\ImportTok{import}\NormalTok{ pyssam}

\NormalTok{tree\_class }\OperatorTok{=}\NormalTok{ pyssam.datasets.Tree(num\_extra\_ends}\OperatorTok{=}\DecValTok{1}\NormalTok{)}
\NormalTok{landmark\_coordinates }\OperatorTok{=}\NormalTok{ np.array(}
\NormalTok{    [tree\_class.make\_tree\_landmarks() }\ControlFlowTok{for}\NormalTok{ i }\KeywordTok{in} \BuiltInTok{range}\NormalTok{(}\DecValTok{0}\NormalTok{, num\_samples)]}
\NormalTok{)}

\NormalTok{ssm\_obj }\OperatorTok{=}\NormalTok{ pyssam.SSM(landmark\_coordinates)}
\NormalTok{ssm\_obj.create\_pca\_model(ssm\_obj.landmarks\_scale)}
\NormalTok{mean\_shape\_columnvector }\OperatorTok{=}\NormalTok{ ssm\_obj.compute\_dataset\_mean()}
\end{Highlighting}
\end{Shaded}

\subsection{Shape and appearance modelling of lung shape and chest X-ray images}

In the following example, we show a real application where 3D landmark
for the left lower lung lobe are projected onto digitally reconstructed
X-rays (Väänänen et al., 2015) and the gray-value is used to obtain
appearance. Example landmark data was obtained using an automatic
algorithm (Ferrarini et al., 2007). Appearance information is extracted
from the X-ray images using \texttt{AppearanceFromXray} (part of
\texttt{pyssam.utils}). We use landmarks, X-ray images as well as origin
and pixel spacing information for the X-ray images to extract appearance
as follows

\begin{Shaded}
\begin{Highlighting}[]
\NormalTok{appearance\_xr }\OperatorTok{=}\NormalTok{ pyssam.AppearanceFromXray(}
\NormalTok{    IMAGE\_DATASET, IMAGE\_ORIGIN, IMAGE\_SPACING}
\NormalTok{)}
\NormalTok{appearance\_values }\OperatorTok{=}\NormalTok{ appearance\_xr.all\_landmark\_density(}
\NormalTok{    landmarks\_coordinates}
\NormalTok{)}
\end{Highlighting}
\end{Shaded}

The SSAM can then be trained in a similar way as the SSM in
\autoref{sec:tree} with the following code snippet:

\begin{Shaded}
\begin{Highlighting}[]
\NormalTok{ssam\_obj }\OperatorTok{=}\NormalTok{ pyssam.SSAM(landmark\_coordinates, appearance\_values)}
\NormalTok{ssam\_obj.create\_pca\_model(ssam\_obj.shape\_appearance\_columns)}
\NormalTok{mean\_shape\_appearance\_columnvector }\OperatorTok{=}\NormalTok{ ssam\_obj.compute\_dataset\_mean()}
\end{Highlighting}
\end{Shaded}

The shape and appearance modes can then be computed based on the model
parameters (\texttt{ssam.model\_parameters}). The computed model
parameters (eigenvectors and eigenvalues of the covariance matrix) can
be used to morph the shape and appearance using
\texttt{ssam.morph\_model} (part of \texttt{StatisticalModelBase} in
\autoref{fig:code}) by

\begin{equation}\label{eq:ssm}
\boldsymbol{x} \approx \bar{\boldsymbol{x}} + \boldsymbol{\Phi} \cdot \boldsymbol{b}  
\end{equation}

where \(\boldsymbol{x}\) is a new array containing shape and appearance,
\(\bar{\boldsymbol{x}}\) is the training dataset mean shape and
appearance, \(\boldsymbol{\Phi}\) is the model principal components
(eigenvectors of the training data covariance matrix),
\(\boldsymbol{b}\) is the model parameters, which is an array of weights
unique to each data sample. The model parameter a mode \(m\) should be
within
\([-3\sqrt{\boldsymbol{\sigma_m^2}}, 3\sqrt{\boldsymbol{\sigma_m^2}}]\),
where \(\sigma_m^2\) is the explained variance of \(m\) (\(m^{th}\)
largest eigenvalue of the covariance matrix) (Cootes et al., 1995).

\begin{figure}
\centering
\includegraphics[width=1\textwidth,height=\textheight]{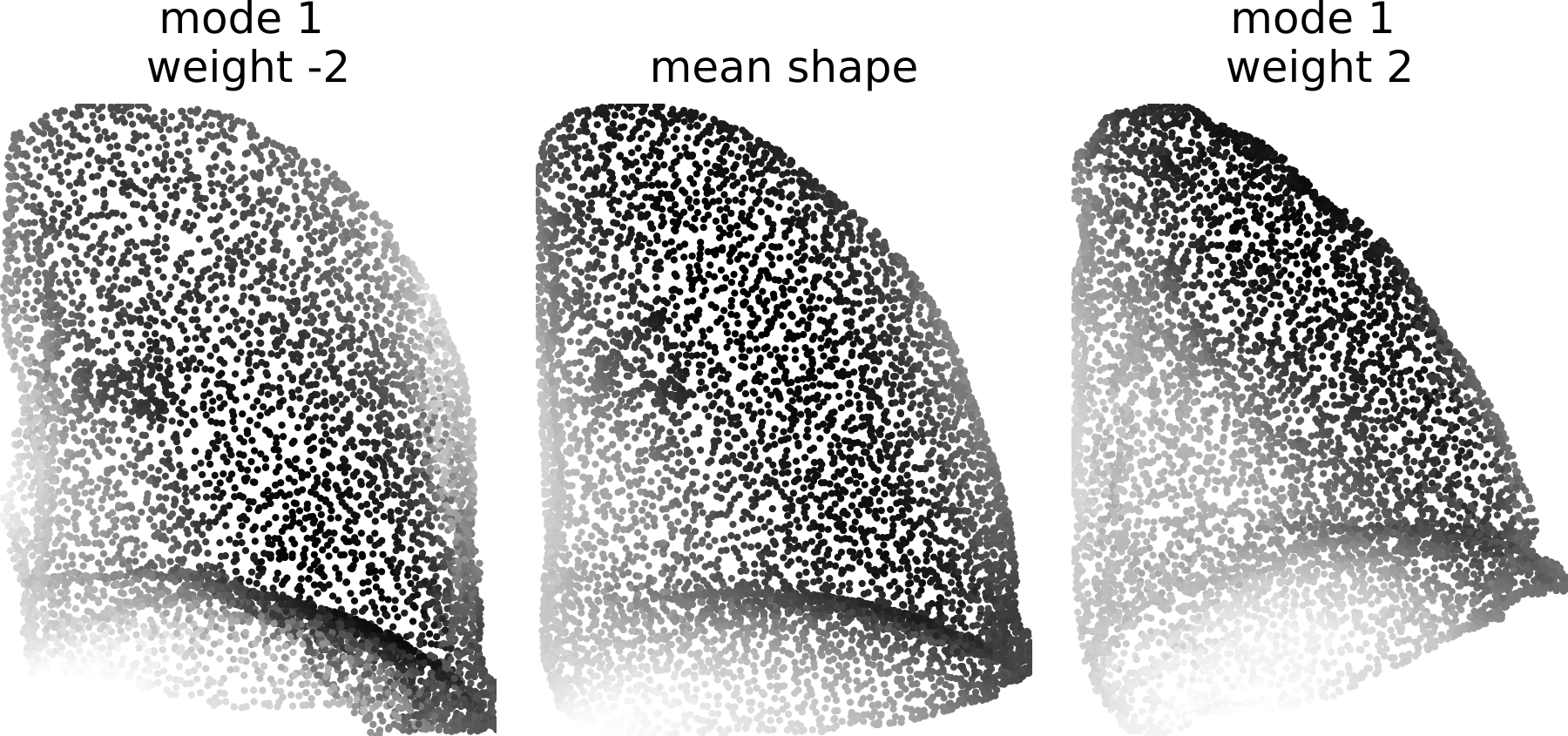}
\caption{First mode of SSAM variation for lung lobe dataset. Panels show
shape and appearance morphed using \texttt{ssam.morph\_model} method and
varying the model parameters (\texttt{ssam.model\_parameters}), from -2,
0 (mean shape) and 2.\label{fig:lungSSAM}}
\end{figure}

Each mode of shape and appearance variation is visualised, as shown for
a representative mode in \autoref{fig:lungSSAM}. This shows how lung
shape influences the gray-value of lung pixels on the X-ray image. In
this case, the change in shape and appearance are mainly due to how the
lung interacts with adjacent structures such as the heart, rib cage and
diaphragm.

\section*{Acknowledgement}

JW was funded by a 2019 PhD Scholarship from the Carnegie-Trust for the
Universities of Scotland.


\hypertarget{references}{%
\section*{References}\label{references}}
\addcontentsline{toc}{section}{References}

\hypertarget{refs}{}
\begin{CSLReferences}{1}{0}
\leavevmode\hypertarget{ref-baka20112D3D}{}%
Baka, N., Kaptein, B. L., Bruijne, M. de, Walsum, T. van, Giphart, J.,
Niessen, W. J., \& Lelieveldt, B. P. (2011). {2D--3D shape
reconstruction of the distal femur from stereo X-ray imaging using
statistical shape models}. \emph{Medical Image Analysis}, \emph{15}(6),
840--850.

\leavevmode\hypertarget{ref-bhalodia2018deepssm}{}%
Bhalodia, R., Elhabian, S. Y., Kavan, L., \& Whitaker, R. T. (2018).
{DeepSSM}: A deep learning framework for statistical shape modeling from
raw images. \emph{International Workshop on Shape in Medical Imaging},
244--257.

\leavevmode\hypertarget{ref-cates2017shapeworks}{}%
Cates, J., Elhabian, S., \& Whitaker, R. (2017). Shapeworks:
Particle-based shape correspondence and visualization software. In
\emph{Statistical shape and deformation analysis} (pp. 257--298).
Elsevier.

\leavevmode\hypertarget{ref-cootes1995active}{}%
Cootes, T. F., Taylor, C. J., Cooper, D. H., \& Graham, J. (1995).
Active shape models-their training and application. \emph{Computer
Vision and Image Understanding}, \emph{61}(1), 38--59.

\leavevmode\hypertarget{ref-ferrarini2007games}{}%
Ferrarini, L., Olofsen, H., Palm, W. M., Van Buchem, M. A., Reiber, J.
H., \& Admiraal-Behloul, F. (2007). GAMEs: Growing and adaptive meshes
for fully automatic shape modeling and analysis. \emph{Medical Image
Analysis}, \emph{11}(3), 302--314.

\leavevmode\hypertarget{ref-heimann2009statistical}{}%
Heimann, T., \& Meinzer, H.-P. (2009). Statistical shape models for {3D}
medical image segmentation: A review. \emph{Medical Image Analysis},
\emph{13}(4), 543--563.

\leavevmode\hypertarget{ref-irving2011segmentation}{}%
Irving, B., Goussard, P., Gie, R., Todd-Pokropek, A., \& Taylor, P.
(2011). Segmentation of obstructed airway branches in {CT} using airway
topology and statistical shape analysis. \emph{2011 IEEE International
Symposium on Biomedical Imaging: From Nano to Macro}, 447--451.

\leavevmode\hypertarget{ref-luthi2012statismo}{}%
Lüthi, M., Blanc, R., Albrecht, T., Gass, T., Goksel, O., Büchler, P.,
Kistler, M., Bousleiman, H., Reyes, M., Cattin, P., \& others. (2012).
Statismo-a framework for {PCA} based statistical models. \emph{The
Insight Journal}, \emph{2012}, 1--18.

\leavevmode\hypertarget{ref-osanlouy2020lung}{}%
Osanlouy, M., Clark, A. R., Kumar, H., King, C., Wilsher, M. L., Milne,
D. G., Whyte, K., Hoffman, E. A., \& Tawhai, M. H. (2020). Lung and
fissure shape is associated with age in healthy never-smoking adults
aged 20--90 years. \emph{Scientific Reports}, \emph{10}(1), 1--13.

\leavevmode\hypertarget{ref-tang2019automatic}{}%
Tang, H., Zhang, C., \& Xie, X. (2019). Automatic pulmonary lobe
segmentation using deep learning. \emph{arXiv Preprint
arXiv:1903.09879}.

\leavevmode\hypertarget{ref-vaananen2015generation}{}%
Väänänen, S. P., Grassi, L., Flivik, G., Jurvelin, J. S., \& Isaksson,
H. (2015). {Generation of 3D shape, density, cortical thickness and
finite element mesh of proximal femur from a DXA image}. \emph{Medical
Image Analysis}, \emph{24}(1), 125--134.

\end{CSLReferences}

\end{document}